\documentclass[aps,twocolumn,floats,prx,nofootinbib]{revtex4}
\usepackage{graphics,graphicx,epsfig}
 \usepackage{textgreek,bm}
\usepackage{amssymb,color}
\usepackage{amsmath,url,hyperref}
\DeclareMathOperator{\sgn}{sgn}
\usepackage{epsf,epstopdf,wrapfig}

\begin{document}

\title{Moving boundaries:  An appreciation of John Hopfield}

\author{William Bialek}
\affiliation{Joseph Henry Laboratories of Physics, Princeton University, Princeton, NJ 08544 USA}

\begin{abstract}
The 2024 Nobel Prize in Physics was awarded to John Hopfield and Geoffrey Hinton, ``for foundational discoveries and inventions that enable machine learning with artificial neural networks.''   As noted by the Nobel committee, their work moved the boundaries of physics.  This is a brief reflection on Hopfield's work, its implications for the emergence of biological physics as a part of physics,  the path from his early papers to the modern revolution in artificial intelligence, and prospects for the future.
\end{abstract}

\date{\today}

\maketitle

\section{Introduction}

I once asked John Hopfield how he decided to change fields from condensed matter physics to biophysics.  ``I didn’t change,'' he said, ``I kept doing the same things and the fields moved past me.''   On October 8, 2024, John was awarded the Nobel Prize in Physics, jointly with Geoffrey Hinton \cite{nobel_website}.  This marks the first time that the physics prize recognizes work in biological physics,  a milestone for our field. The emergence of biological physics as a proper branch of physics is a long and complicated story \cite{decadal_survey}.  But, more than any other single individual, John saw how a theoretical physicist could engage with the beautiful and sometimes mysterious phenomena of life.  His theories of neural networks  \cite{hopfield_1982,hopfield_1984}, recognized by the Nobel committee,  are only a small part of this.

Hinton's path was different,  having started as a student of experimental psychology and then artificial intelligence (AI), which of course meant something very different in the 1970s.  His PhD adviser Christopher Longuet--Higgins was a theoretical chemist, and a very physical chemist, having discovered (among other things) a specific version of what we now call geometric phases in quantum mechanics \cite{longuet-higgins+al_1958}.    The first of Hinton's many accomplishments cited by the Nobel committee is the Boltzmann machine \cite{ackley+al_1985}, which drew (as the name suggests!) from ideas in statistical physics, and built directly on Hopfield's work.

The intertwining of physics and AI reaches back  to the foundational papers on perceptrons, some of which appeared in {\em Reviews of Modern Physics} in the early 1960s \cite{block_1962,block+al_1962}.  Artificial intelligence has obvious origins in efforts to emulate life, but it was never clear if there were principles  to be learned from real brains that would transfer to the engineering context.  After all, airplanes don't flap their wings, but they do obey the same equations of fluid mechanics that govern flight in birds, and as a result many of the principles of lift generation are the same.

One approach to AI was to give abstract formulations of the problems solved by real brains, and then to develop hand crafted algorithms that address these abstract problems.  Hopfield and Hinton presented a very different strategy: writing models for the dynamics of interacting neuron--like elements and asking what computational functions can emerge from these dynamics.  Certainly for Hopfield, ``emerge'' meant emerge in the sense that other collective phenomena emerge in many--body systems---the rigidity of solids, the polarization of magnets, and more.  These are problems for which statistical mechanics gives us a natural language.

Great ideas have multiple origin stories.  Often these stories are seen as competing, and the greater the idea's impact the more seems to be at stake, deciding which discipline or whose intellectual ancestors should get credit for what happened.  Perhaps it is more interesting that the multiple stories can all be true:  different individuals and even different communities come to more or less the same realizations by different paths. 

For the biological physics community it matters that one of the paths to the current AI revolution went through efforts to gain a physicist's understanding of collective dynamics in the brain.  In looking at Hopfield's work, one can see this as part of a larger project to gain a physicist's understanding of life.  This is not to say that Hopfield had a program to work in sequence on the phenomena of life at different scales of organization, from single molecules up to brains. But this is what happened.

The Nobel Prize provides the occasion to take a brief tour of Hopfield's broader contributions to the emergence of biological physics, the roots of his thinking in earlier work on condensed matter physics, and the explosion of work that followed his ideas about neural networks.   To keep things compact, most technical matters will be suppressed, but I hope to capture something of the intellectual style and provide some pointers to the future.

\section{Before neural networks}

Scientific careers are not deterministic trajectories.  We should avoid the temptation of thinking that the advances recognized by the Nobel Prize ``grew out'' of Hopfield's earlier work.  At the same time, there is a remarkable intellectual coherence across multiple topics---and across eight decades, from the 1950s to the 2020s.

John's intellectual adventures began with the seemingly modest problem of calculating the dielectric behavior of insulating crystals \cite{hopfield_1958}.  
For a single electron to absorb energy it must make a transition from the (filled) valence band to the (empty) conduction band, leaving behind a hole. The electron and hole can form a bound state, called an exciton.  John's contribution was to show that the compounding of excitations did not end there:  as light passes through the crystal, photons couple to excitons so that the independent and long--lived excitations are a mixture, which he termed polaritons.  Polaritons are bosons, and it is extraordinary that fifty years after Hopfield's work there were observations indicating that these excitations exhibit Bose--Einstein condensation and superfluidity, thus forming an essentially non--equilibrium quantum fluid \cite{amo+al_2009}.  We are by now accustomed to the idea that the independent, long--lived ``modes'' or elementary excitations of a system can be very different from its microscopic constituents, but this was far from obvious in the late 1950s.

\begin{figure}
\includegraphics[width=\linewidth]{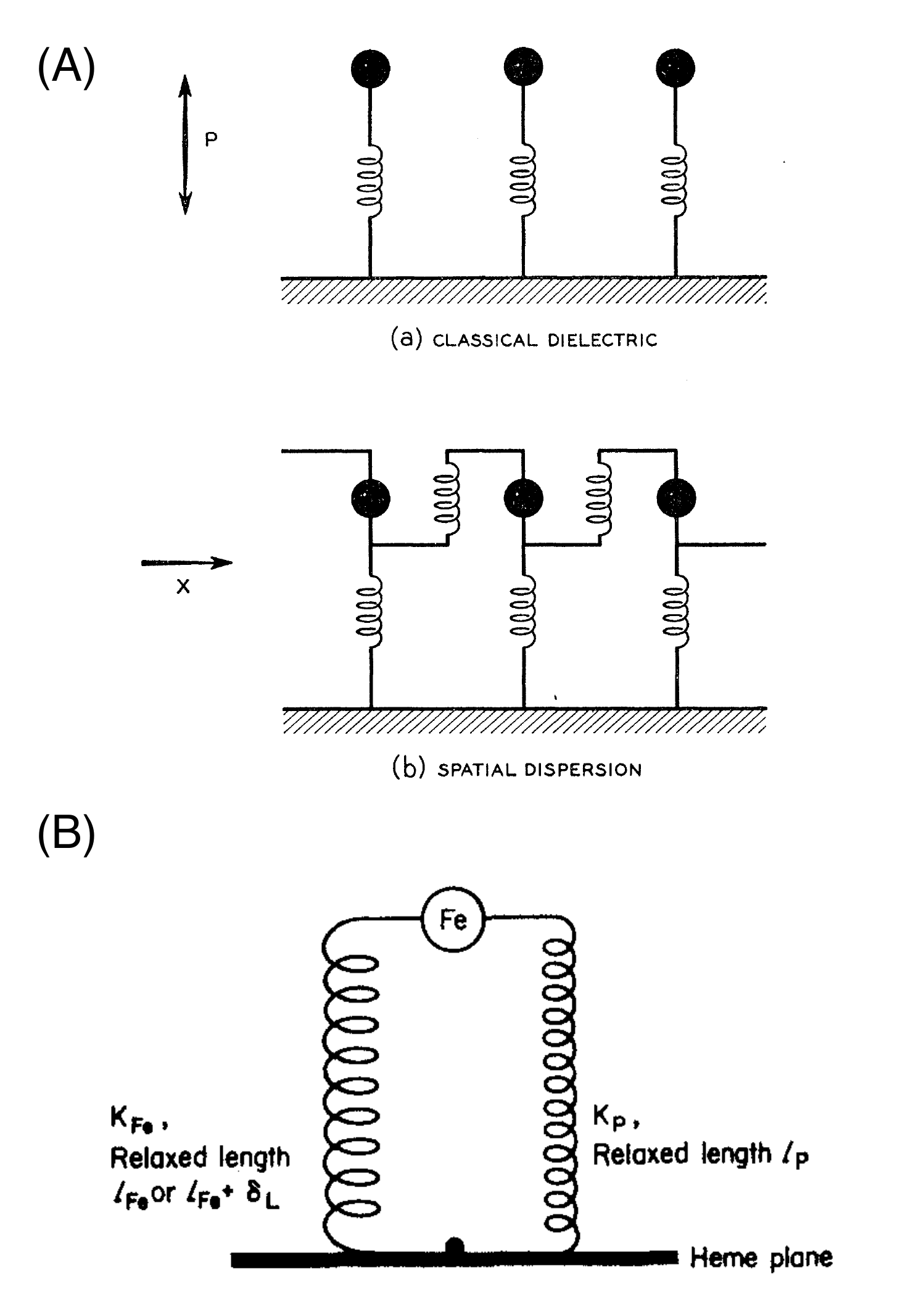}
\caption{Simple models. (A) In a classical dielectric, localized charges oscillate in response to an applied field.  To generate spatial dispersion, there must be a path for energy flow, schematized by (effective) lateral springs connecting the charges \cite{hopfield+thomas_1963}.  (B) In hemoglobin, the iron atom is bound to the heme group by spring of stiffness $K_{\rm Fe}$ that changes its equilibrium length ($\ell_{\rm Fe} \rightarrow \ell_{\rm Fe} + \delta_L$) when a ligand such as oxygen binds to the iron.  In addition the iron atom is held by the protein as whole, schematized as a spring of stiffness $K_{\rm P}$ with a different equilibrium length $\ell_{\rm P}$   \cite{hopfield_1973}. \label{springs}}
\end{figure}

The interaction of light and matter continued as a theme for more than a decade.  In particular, Hopfield had an extended theory/experiment collaboration with  David Thomas.  Classical non--magnetic insulators  are characterized by a local dielectric response, so that the polarization within a  sub--wavelength volume depends only on the electric field in that volume.  Thomas and others showed that semiconductors exhibited behaviors that could not be accounted for in this way.  John emphasized that non--locality must mean that there is another path for energy transport through the system, schematized in startlingly simple terms  in Fig \ref{springs}A.  Theory and experiment came together in a landmark paper \cite{hopfield+thomas_1963}, and Hopfield and Thomas went on to share the 1969 Oliver Buckley Prize for Condensed Matter Physics from the American Physical Society. 

Hopfield's first foray into the physics of life explored the cooperativity of oxygen binding to hemoglobin.  The phenomenology went back at least to AV Hill's work in 1910, from which we get the ubiquitous ``Hill function;'' hemoglobin and myoglobin were the first proteins whose structures were determined through the analysis of X--ray diffraction experiments \cite{perutz+al_1960,kendrew+al_1960}. Inspired in part by these structures, J Monod, J Wyman, and J--P Changeux proposed a model  in which the binding of individual molecules  shifts the equilibrium between two different protein structures \cite{monod+al_1965}.  To be thermodynamically consistent the different structures must have different binding energies for the individual molecules, and this generates an effective cooperativity between events at distant sites---allostery.  If this MWC model is correct, one should see the switching between protein structures as part of the kinetics of oxygen binding or dissociation, and Hopfield worked with his experimental colleagues RG Shulman and Seiji Ogawa to show that this works in great detail \cite{hopfield+al_1971}. They could push further and show that the binding of each oxygen molecule was independent so long as the protein remained in one state or the other \cite{shulman+al_1972}: cooperativity arises only when we average over the fluctuations or switching between protein states.

Hopfield's thinking about hemoglobin culminated in a model where the energies that drive functional changes are distributed throughout the molecule rather than localized in particular chemical bonds, as schematized in Fig \ref{springs}B  \cite{hopfield_1973};  I could not resist the juxtaposition of this schematic with the one from a decade before.  This view  was completely novel and far ahead of its time. 
In his own words: ``In the simplest distributed model ...  The rest of the molecule will be described by a second spring connecting the same two points, but having a different equilibrium length. This spring represents the connection between the central region of the heme group and the iron atom via the path leading to the outer edge of the heme, through the protein, and ultimately to the proximal histidine.'' All the complexity of protein structure and flexibility is summarized by an effective  spring.  Stretching the spring  also shifts the electronic energy levels of the heme group, and this leads to quantitative predictions for the shape of the optical absorption spectrum and its shift as the protein switches between different states \cite{hopfield_1973}.  Hopfield's thinking about hemoglobin thus combined radical simplification and engagement with experimental detail, a pattern that would repeat. 

Next, Hopfield explored the  interplay between classical and quantum dynamics in biological electron transfer,  a process that operates to secure the energy supply of all cells \cite{hopfield_1974a}.  Experiments demonstrating the persistence of these reactions at very low temperatures had suggested that quantum tunneling was relevant, rekindling dreams of many physicists for a role of quantum mechanics in biology.  Hopfield presented a theory which clarified the rampant confusion: electrons almost always tunnel, but the accompanying structural changes in the protein can be classical at high temperatures and quantum mechanical at low temperatures.  As with hemoglobin, the protein was approximated as harmonic, not because there are no nonlinearities but because the force on any single mode of the large molecule is small.  While the energetics of hemoglobin probed only the stiffness of the effective springs, the temperature dependence of reaction rates probes the dynamics as the thermal energy crosses the energies of vibrational quanta.  Once again, functional behaviors are connected to spectroscopic signatures, including the prediction of new absorption bands  \cite{hopfield_1977,goldstein+bearden_1984}.

The interplay between classical and quantum dynamics identified by Hopfield would become even more important as it became clear that the rapid initial steps of photosynthetic electron transfer occur with essentially temperature independent rates even at room temperature.  His work focused attention on the problem of calculating the quantum mechanical matrix elements for electron hopping from one site to another, a problem which John addressed with his students David Beratan and Jos\'e Nelson Onuchic; their results substantially revised our views about the ``pathways'' for electron transfer  through proteins \cite{beratan+al_1987} and inspired a generation of experiments. 

An astonishing two months after his work on electron transfer, Hopfield emphasized that many  processes crucial to  life---from the replication of DNA to the synthesis of proteins---involve discrimination among subtly different molecular components, and that  these discriminations occur with error rates much smaller than expected  if the reactions responsible for selectivity are allowed to come to thermal equilibrium.  In general, he argued that greater accuracy could be achieved if cells expend energy to hold the system away from equilibrium, at which point ``kinetic proofreading'' or error--correction  becomes possible; he could show that  the outlines of such a scheme were consistent with available data on several systems.  While earlier experiments had shown, for example, that the enzymes involved in DNA replication were capable of `backing up' to remove incorporated nucleotides, the fundamental connection between genuine error correction and the dissipation of energy was made by Hopfield \cite{hopfield_1974b}.  In effect, cells build Maxwell demons to sort molecules.
  
Typical of John's scientific style, he  set out to collaborate with Tetsuo Yamane on experiments that provided the first strong evidence for proofreading \cite{yamane+hopfield_1977},  and went on to present variations on the theme in which the non--equilibrium drive is more distant from the essential synthetic steps \cite{hopfield_1980},  showing how the crucial physics could be missed in the conventional biological focus on ``mechanism.''  Hopfield's work on proofreading has strong echoes today, nearly fifty years later, in ideas about the thermodynamics of computation and information transmission, in the analysis of fluctuation relations and thermodynamic uncertainty, as well as in discussions of precision in a wide range of biological processes from the immune response to the topological transitions of DNA.

It is difficult to overstate the importance of proofreading.  If you ask most people why genetic information is transmitted reliably, they will say that it is because of base pairing---A pairs with T and C with G because the structures are complimentary; this is the primordial example of structure determining function in biological molecules.  But this misses the quantitative facts by such a large margin that it seems qualitatively wrong.  The free energy differences between correct and incorrect base pairs are sufficient to drive error probabilities down to $\sim 10^{-4}$ if the reactions come to equilibrium, but the observed error probabilities are $\sim 10^{-8}$.  The conventional answer thus is only the square root of the story, and the difference is kinetic proofreading.  Without proofreading, almost every one of our genes would carry at least one mutation relative to the sequences provided by our parents, and almost every protein molecule in every cell would have at least one incorrect amino acid.

\section{Neural networks}

Neurons are complicated objects.  Their structural complexity had been revealed by Cajal by 1900; over the course of the twentieth century, generations of experiments revealed the complexity of their electrical dynamics and the diversity of molecules that generate these behaviors.  Parallel to these developments---which culminate in detailed mathematical descriptions of the molecular dynamics of ion channels, synapses, and receptors---there have been persistent efforts to write very simplified models for single neurons and use these models to understand what emerges as neurons are connected into networks.  

The search for simplified models goes back at least to 1943, with  McCulloch and Pitts \cite{mcculloch+pitts_1943}.  They abstracted from the ``all or none'' character of the action potential to describe each neuron as active ($\sigma_{\rm i} =+1$) or inactive ($\sigma_{\rm i} =-1$), with ${\rm i} = 1,\, 2,\, \cdots ,\, N$ indexing the $N$ cells in the network.  Dynamics were  pictured as occurring in discrete time, with each neuron taking a weighted sum of inputs from other neurons and comparing with a threshold to decide if it would be active or inactive at the next step:
\begin{equation}
\sigma_{\rm i}(t+1) = \sgn\left[\sum_{{\rm j}\neq {\rm i}} J_{\rm ij} \sigma_{\rm j} (t) - \theta_{\rm i}\right] .
\label{dynamics1}
\end{equation}
The weights $J_{\rm ij}$ are attached to the connections or synapses between cells, and what the network can compute is controlled by these weights.  The $\sigma_{\rm i} = \pm 1$ variables remind us of an Ising model, and by the 1970s WA Little, GL Shaw, and others were making this connection with statistical mechanics more explicit \cite{little_1974,little+shaw_1975}.  In 1973, LN Cooper discussed the way in which memories could be imprinted in synaptic weights at a Nobel Symposium on Collective Properties of Physical Systems \cite{cooper_1973},\footnote{This was a remarkable conference \cite{NobelSymposium}.  KG Wilson spoke about the renormalization group and the Kondo effect; RC Richardson and DM Lee described the discovery of superfluidity in $^3$He, PW Anderson and AJ Leggett discussed the theory of these new phases, and PG de Gennes described anisotropic superfluids more generally; LP Gor'kov discussed superconductivity in low dimensional systems and AJ Heeger described a one--dimensional organic metal;  W Kohn discussed metal surfaces; and Hopfield spoke about hemoglobin.  There are many other papers worth reading; this list includes only the ten  speakers who received a Nobel Prize in the years after the conference. It is important that all these problems were seen as part of one subject, collective phenomena, and that problems related to the physics of life had a place on this last, as emphasized by Anderson in his conference summary.}  emphasizing that neural networks are a physics problem. Despite these and other precursors, Hopfield's papers in 1982 and 1984 came as a surprise.

\begin{figure}
\includegraphics[width=\linewidth]{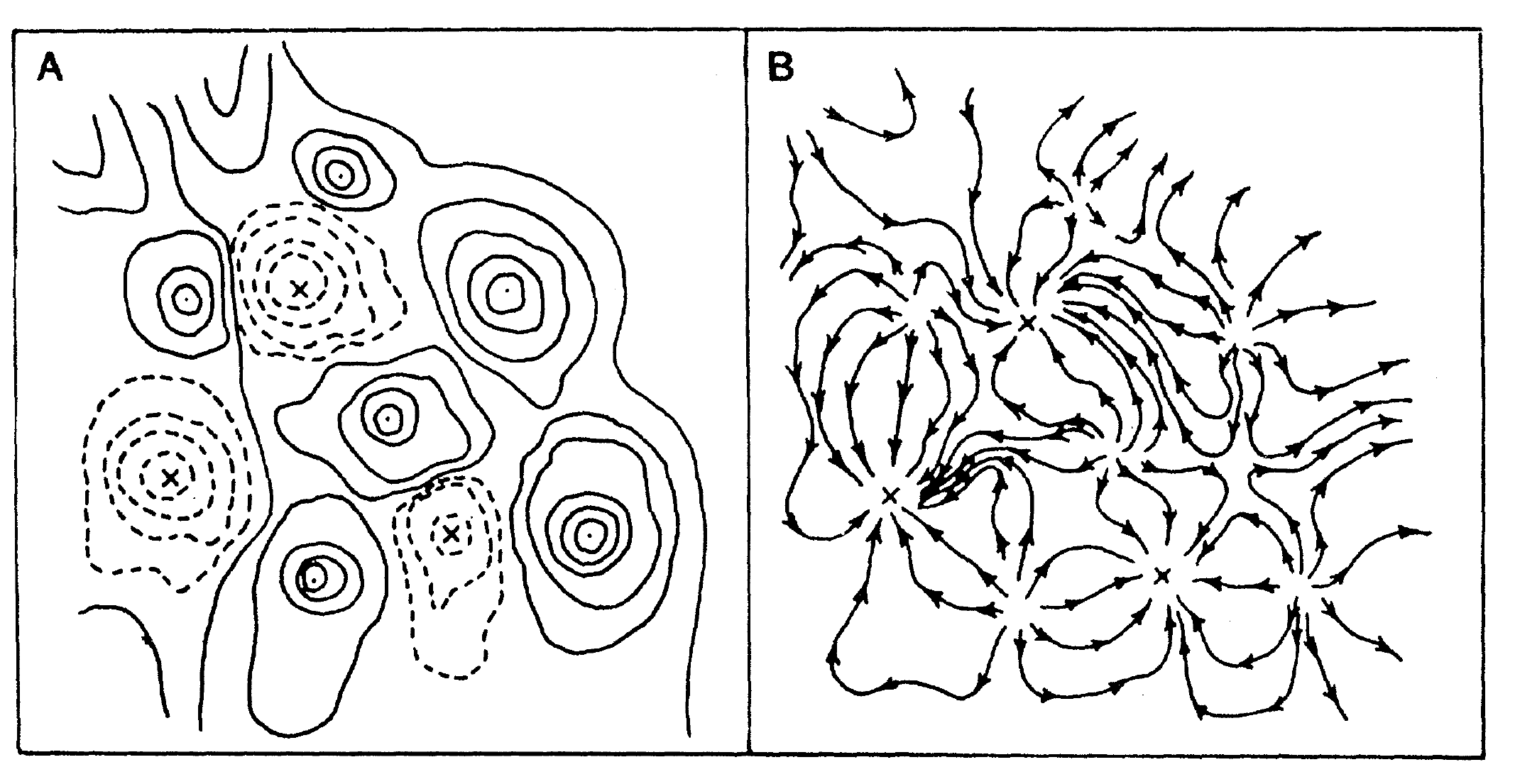}
\caption{Energy landscape and trajectories in a model of neural networks \cite{hopfield+tank_1986}.  (A) Solid contours are above a mean level and dashed contours below, with X marking fixed points at the bottoms of energy valleys.  (B) Corresponding dynamics, shown as a flow field.  
\label{landscape}}
\end{figure}

Hopfield made progress by taking a step backward. Instead of considering general matrices $J_{\rm ij}$, he focused on symmetric synapses, so that $J_{\rm ij} = J_{\rm ji}$ \cite{hopfield_1982}.  With this assumption, the dynamics in Eq (\ref{dynamics1}) has a Lyapunov function, so we can envision the network as sliding downhill on a landscape, and it is natural to think of this landscape as an effective energy function (Fig \ref{landscape}).  This picture is even clearer in generalizations of the model where the activity of each neuron becomes a continuous variable \cite{hopfield_1984}. Hopfield emphasized that coming to a rest at the minimum of the energy is a computation, analogous to recalling a memory.  More generally he drew attention to the fact that computation is dynamical, following a trajectory from inputs (including a program, in conventional computers) to a final state that represents the answer.  As in classical mechanics, one can then gain insight by asking about global properties of the trajectories, many of which are stable against variations in the local rules.

In the simple case where the thresholds $\theta_{\rm i} = 0$, the energy function is
\begin{equation}
E\left(  \bm{\sigma} \right)= - {1\over 2} \sum_{\rm ij} J_{\rm ij}\sigma_{\rm i}\sigma_{\rm j} ,
\label{HM1}
\end{equation}
where $\bm{\sigma} = \{\sigma_{\rm i}\}$ is shorthand for the state of the entire network.
This is an Ising model, with pairwise interactions among the ``spins'' $\sigma_{\rm i}$.  Importantly is possible to ``program'' the network so that the stable final states of the network are close to some specific stored patterns.  This is done by choosing
\begin{equation}
J_{\rm ij}= J\sum_{\mu =1}^K \xi_{\rm i}^\mu \xi_{\rm j}^\mu ,
\label{HM2}
\end{equation}
where the $\xi_{\rm i}^\mu = \pm 1$ are the $K$ binary patterns we'd like to store.   If $K=1$ we can see that the energy function is gauge equivalent to the mean--field ferromagnet, and so the ground state is  $\bm{\sigma} = \bm{\xi}^1$.   If $K$ is not too large, and the patterns $\xi_{\rm i}^\mu$ are sufficiently random, then there will be multiple ground states at $\bm{\sigma} = \bm{\xi}^1$, $\bm{\sigma} = \bm{\xi}^2$, ... , $\bm{\sigma} = \bm{\xi}^K$.  In his original work Hopfield gave a rough estimate that this picture of multiple energy minima at each of the stored patterns should keep working until $K\sim 0.15 N$ \cite{hopfield_1982}.    Not long before Hopfield's work there had been dramatic developments in the statistical mechanics of disordered systems \cite{mezard+al_1987}, and soon these tools were brought to bear, making the picture precise. If we define the overlap between the network state and  the stored patterns, 
\begin{equation}
m_\mu = {1\over N}\sum_{{\rm i}=1}^N \sigma_{\rm i}\xi_{\rm i}^\mu ,
\end{equation}
then  for $K < \alpha_c N$, the mean $\langle m_\mu\rangle$
is close to one for some particular $\mu$ and not the others, while for $K > \alpha_c N$ we find $m_\mu \sim 0$ \cite{amit+al_1985,amit+al_1987}; this is a genuine phase transition as $N\rightarrow\infty$.  Thus in the Hopfield model the successful retrieval of memories is an emergent phenomenon in the same sense that magnetism in an emergent phenomenon.

If the system is in a state $\bm{\sigma}^t$ at some moment in time, and we would like to add this to list of stored patterns, then the synaptic strengths should be adjusted as
\begin{equation}
J_{\rm ij} \rightarrow J _{\rm ij} + J \sigma_{\rm i}^t \sigma_{\rm j}^t .
\label{hebb}
\end{equation}
We notice that this is a local learning rule:  what happens at the synapse between neurons $\rm i$ and $\rm j$ depends only on the states of those two neurons, and not on the rest of the network.  This is surprising because ground states are a property of the network as a whole, yet they can be programmed without global knowledge.  

In the Hopfield model memories are stable patterns---attractors, to use the language of dynamical systems---such that the activity of any one cell is driven by all the other neurons in the network, self--consistently; this 
provided a mathematically precise version  of ``reverberation'' in neural circuits.  The learning rule in Eq (\ref{hebb})  means that synapses are strengthened between neurons that are simultaneously active.  Summarized as ``fire together, wire together,'' this conception of synaptic plasticity goes back to Donald Hebb in the 1940s \cite{hebb_1949} and has precursors in William James's writings from the turn of the twentieth century \cite{james_1904}.  Direct observations of synaptic plasticity, roughly conforming to the Hebbian Eq (\ref{hebb}), came not long before Hopfield's papers \cite{bliss+lomo_1973}.\footnote{Some subtleties: James actually said (in our notation) that the connection $J_{\rm ij}$ should be strengthened if activity in neuron $\rm j$ {\em caused} activity in neuron $\rm i$; this version of the idea would reappear with the discovery of plasticity rules that have a precise dependence on the timing of action potentials in the two cells \cite{caporale+dan_2008}.  With thresholds set to zero, Eq (\ref{hebb}) predicts that $J_{\rm ij}$ will be increased if activity in cells $\rm i$ and $\rm j$ are correlated, which is subtly different than being active together.  This importance of this distinction was emphasized by Terry  Sejnowski  in his 1978 PhD thesis \cite{sejnowski_1978}; Hopfield was his adviser.}  Hopfield's work   thus provided a framework in which many ideas about neural circuits fit together into a coherent theory.

Hopfield's model showed how the dynamics in a network of neurons could be understood as minimizing an energy function.  Beyond memory retrieval,  many other computational problems can be phrased in this variational form.  Central issues in the theory of computational complexity are illustrated by such optimization problems, perhaps most famously the ``traveling salesman'' who has to visit each city on his route exactly once while taking the shortest possible path.  Hopfield and David Tank constructed a neural circuit in which the effective energy function is essentially the cost function of the salesman's path, so that the pattern of activity in a stable state of the network represents a solution to the problem \cite{hopfield+tank_1985}.  This opened a path to thinking more generally about how neural circuit dynamics becomes functional, computational dynamics in the brain  \cite{hopfield+tank_1986,tank+hopfield_1987}.

I remember Edward Witten once starting a seminar by saying that he was trying to understand how string theory could recover certain stylized facts about the world.  ``Stylized facts'' is a wonderful description of John’s connection to biology in his classic neural network papers.  Only someone with intimate knowledge of the facts could choose so well which ones to use in guiding his thinking, and which to set aside.  The wisdom of his choices is borne out by subsequent events.

\section{What happened after}

It can be hard to say exactly why particular developments have a huge impact on the scientific community.  In the case of the Hopfield papers, part of the answer is that he connected many different communities, encouraging people to cross boundaries.\footnote{References in this section are illustrative rather than exhaustive, and should not be interpreted as assignments of credit for particular developments.  The crucial papers \cite{hopfield_1982,hopfield_1984,hopfield+tank_1985,hopfield+tank_1986} have accumulated more  than 38,000 citations, so any listing necessarily is a sparse sampling of their impact.}

Seen from a more traditional physics perspective, the Hopfield model in Eqs (\ref{HM1}, \ref{HM2}) is a very particular kind of Ising spin glass, and this is an especially useful point of view if we imagine that the patterns ${\bm \xi}^\mu$ that we are trying to store are chosen at random. Hopfield's work came just after Giorgio Parisi's solution of the mean--field spin glass \cite{mezard+al_1987}, which introduced new tools for a large class of statistical mechanics problems.  It quickly became clear that neural networks are a rich source of such problems.  One legacy of this is in the category ``disordered systems and neural networks'' that one finds on the physics e--print ar$\chi$iv---a literal redrawing of the community's map of the intellectual landscape.

The Hopfield model emphasized that if one is solving the computational problem of minimizing a function it is useful to think of the function as an energy surface and the computation as dynamics on this surface.  While the original model was deterministic, it is natural to generalize to stochastic dynamics or Brownian motion on the energy surface.  This is what happens in optimization by simulated annealing \cite{kirkpatrick+al_1983}, which emerged almost simultaneously with Hopfield's work.   In the same spirit, several groups began to use analytic tools of statistical physics to explore problems originally formulated as computer science problems \cite{fu+anderson_1986}.  This led, for example, to the discovery of phase transitions in the parameter space of NP--complete problems, and to insights about the origins of computational complexity \cite{monasson+al_1999}.

Given our finite experience, we often remain uncertain about the precise rules that underlie our observations.  This probabilistic character of inference suggested a formulation of learning as a statistical mechanics problem, and neural networks---both the Hopfield model and the feed--forward perceptron models---provided concrete examples of this idea \cite{gardner_1988,levin+al_1990,watkin+al_1993}.  Statistical mechanics methods focus on average or typical behaviors, while the computer science literature on learning theory had emphasized worst case bounds; reconciliation of these approaches produced a compelling thermodynamic picture of learning as a competition between goodness of fit to the data (defining an effective energy) and entropy in the space of models \cite{haussler+al_1996}. This thermodynamic view also allows us to understand how there can be phase transitions in learning \cite{seung+al_1992}, perhaps analogous to the familiar ``aha'' phenomenon.

Central to Hopfield's work was seeing computation as dynamics.  To insure reliability, these dynamics must be insulated from noise.  In a digital computer, naturally analog voltages are restored to discrete values at every tick of the clock.   The picture of dynamics as being downhill on an energy landscape emphasizes that restoration can happen globally, so that trajectories lead to the right answer even if there are small perturbations along the way.  This led to a revival of interest in analog as opposed to digital computation, and to a whole new field of ``neuromorphic'' computing that has produced useful devices with very low power consumption \cite{mead_1990}.

The Hopfield model retrieves and holds discrete memories.  But  the brain can hold memory of a continuous variable; examples range from navigation \cite{treves+al_1992} to the integration of the movement signals in our vestibular system to guide eye movements \cite{seung_1996}.  These functions motivate extensions in which the effective energy landscape has attractors that form a manifold---a line, a ring, a sheet.  These generalizations brought direct connections to experiments on the oculomotor system \cite{major+al_2004a,major+al_2004b}, mammalian grid cells \cite{yoon+al_2013}, and most recently the central complex of flies in which neurons not only seem to have the dynamics expected from a ring attractor but actually have the structure of a ring \cite{seelig+jayaraman_2015}.

To round out the connections to neuroscience, we should look also at some of Hopfield's own work.  Although his original model described neurons with discrete on/off activity,  ultimately motivated by the discreteness of action potentials, the community has usually taken the variables in the Hopfield model and its generalizations to be more coarse descriptions of activity, such as rates of action potential generation averaged over substantial intervals of time.  Hopfield himself  emphasized the role of action potential timing in analog computation \cite{hopfield_1995}, especially in olfaction where the problem of defining and identifying objects is presented in perhaps a purer form than the more familiar examples from vision \cite{hopfield_1991,hopfield_1999}.  With his former student Carlos Brody, John went on to discuss how timing and synchronization can lead to new computational primitives and help us understand the representation of a moment \cite{hopfield+brody_2001}.  With Andreas Herz he explored attractors in the dynamics of networks where the discreteness of spiking is relevant \cite{hopfield+herz_1995,herz+hopfield_1995}.

The first step from the Hopfield model toward modern artificial intelligence was taken by Hinton, together with David Ackley and Terrence Sejnowksi:  the ``Boltzmann machine'' \cite{ackley+al_1985}.  There were several ideas.  The first step is to replace the deterministic dynamics of Eq (\ref{dynamics1}) with a noisy or probabilistic update rule chosen so that after a long time the states of the network are drawn from the Boltzmann distribution
\begin{equation}
P\left(\{\sigma_{\rm i}\}\right) = {1\over Z} \exp\left[ - E(\bm{\sigma})/T\right],
\end{equation}
 where the ``temperature'' $T$ measure the strength of the noise, and the energy $E(\bm{\sigma})$ is from Eq (\ref{HM1}). The goal is to find synaptic connections $J_{\rm ij}$ such that this distribution of states in the network matches the distribution of signals in the outside world.   This has a clear connection to simulated annealing, and also to contemporaneous work in the applied mathematics literature that related statistical physics  to problems of Bayesian inference, especially image interpretation \cite{geman+geman_1984}. 
 
It is impossible to match an arbitrary distribution using only pairwise interactions. The second and crucial idea was  to have the number of neurons be larger than the dimensionality of the signal, so that some neurons are directly connected to the outside world and others are hidden.\footnote{The idea that complex distributions can be represented by pairwise interactions with hidden units is not at all obvious.  Perhaps this should have led people to ask whether the distributions of simultaneous activity in networks of real neurons could be described by such pairwise models.  We now know that this does work, sometimes in surprising quantitative detail \cite{meshulam+bialek_2024}.  As this approach was first being developed, neither its advocates nor its critics made much out of the connection to Boltzmann machines.  In 2006 Sejnowski shared with me his brief email exchange with Hinton, expressing their delight that real brains might be Boltzmann machines after all, although this was a bit of an extrapolation from what was known  at the time \cite{schneidman+al_2006}.}   One can then run the network either with input neurons clamped in the states determined by the input themselves or let the network run freely.  The third idea is to compare the probability $p_{\rm ij}$ that neurons $\rm i$ and $\rm j$ are active simultaneously in these two different configurations, and changes the synaptic connections to bring these two distributions closer to one another.  The result is a learning rule
 \begin{equation}
\Delta J_{\rm ij} \propto -\left(p_{\rm ij}^{\rm free} - p_{\rm ij}^{\rm clamped}\right) .
\end{equation}
Ackley, Hinton, and Sejnowski were able to show that the Boltzmann machine could learn to solve a number of what we now would call small problems.  They emphasized that in this model the structure of the input data  is represented in a way that is distributed across all the hidden units, rather than necessarily having  individual neurons responsible for discrete features of the data.

Such ``distributed representations'' had been discussed, but the standard objection was that they lack the modularity of the more discrete feature decompositions, and thus it would be hard to build such a system.  This view is based on the very successful construction of many engineered systems, including computers:  we have components with assigned functions, these functions persist when the components are connected, and the connections are designed to embody a decomposition of larger goals into component parts.  Neural networks are not like this at all.  The shift from programmed decomposition of problems to a learned, distributed representation was central to subsequent progress.  

\begin{figure}
\includegraphics[width = \linewidth]{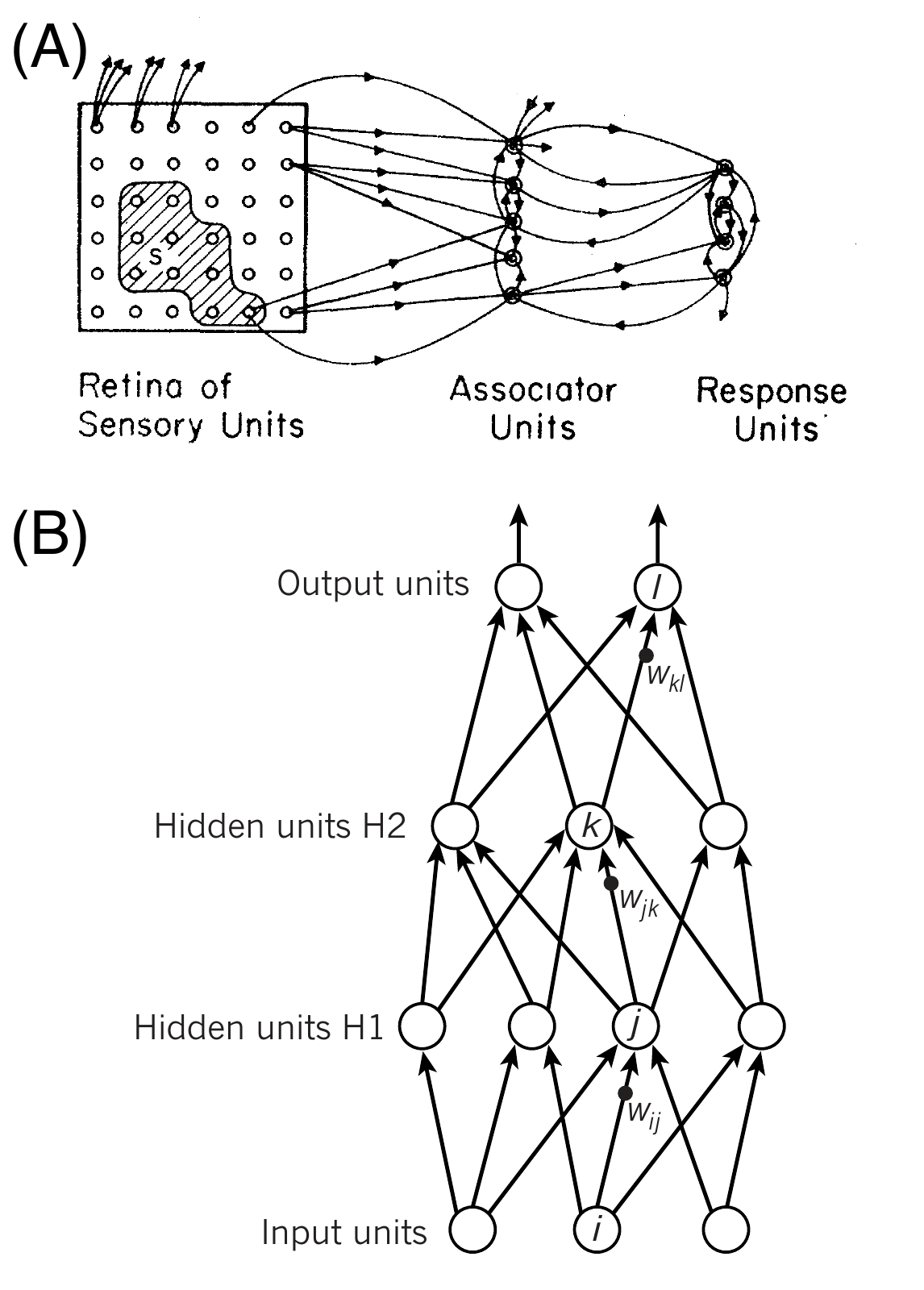}
\caption{Neural networks with a feed--forward architecture, or ``perceptrons.'' (A) An early version from $\sim$1960 \cite{block_1962}.  (B) A modern version  \cite{lecun+al_2015}.  The first steps in the modern AI revolution involved similar networks, with many hidden layers, that achieved human--level performance on image classification and other tasks.  \label{perceptrons}}
\end{figure}

The general model outlined by McCulloch and Pitts is intractable.  Hopfield made progress by restricting his attention to networks with symmetric connections, $J_{\rm ij} = J_{\rm ji}$.  As I hope this discussion has made clear, this simplification resulted not only in technical progress but also in new concepts that had a life outside the confines of a specific model.  The opposite simplification had already been considered (long before) is the perceptron \cite{block_1962,block+al_1962}: a feed--forward architecture in which if $J_{\rm ij}\neq 0$ we are guaranteed that $J_{\rm ji} = 0$, as schematized in Fig \ref{perceptrons}A.  In the 1960s it already was clear that interesting computations would require networks with multiple layers.  But in a feed--forward system information flows from input to output.  If we have access only to inputs and outputs---e.g., images and their correct names---it is not clear how to set the connections among hidden units to optimize the overall network performance.  With David Rumelhardt and Roland Williams, Hinton (re)discovered a startlingly simple solution to this problem \cite{rumelhardt+al_1986}:  we can differentiate the output of the network with respect to internal connection strengths by applying the chain rule, and ``back--propagate'' errors to update those otherwise inaccessible parameters.\footnote{Although we don't usually think of Hopfield in connection with feed--forward networks, there is a fascinating paper with many of his Bell Labs colleagues showing how to use information theoretic measures of performance for these networks and asking whether automatic learning reaches human--like solutions \cite{denker+al_1987}.  This comes just after back--propagation, and gives some sense for the excitement of the times.}

Very quickly, Yann LeCun and his collaborators used back--propagation to train multilayer networks to reach human level performance at reading handwritten zip codes \cite{lecun+al_1989}.  As these ideas developed, an important simplification was that in image processing one expects at least an approximate translation invariance, which was built in through a ``convolutional'' architecture \cite{lecun+al_1998}.

Intermediate between feed--forward networks and the Boltzmann machine are ``restricted Boltzmann machines'' introduced by Paul Smolensky \cite{smolensky_1986}.  The idea is to allow connections between input neurons and the hidden units, but no interactions among either the hidden units or the input units themselves.  This still functions as a model for the probability distribution of the input states, and as applied to image recognition it builds a model of possible images rather than assigning them labels---unsupervised rather than supervised learning.  Hinton developed new algorithms for training these models \cite{hinton_2002}, and with colleagues showed that unsupervised pre--training of individual layers in multilayer feed--forward networks accelerated learning of supervised tasks \cite{hinton+al_2006}.  With the realization that neural networks are in the ``embarrassingly parallel'' class of problems for which graphics processor units are so well suited, the stage was set to train truly deep networks with tens or even hundreds of layers (extensions of Fig \ref{perceptrons}B), reaching human level performance on complex image classification tasks, and more \cite{lecun+al_2015}.

Modern networks are trained by very different methods, often with layers of practical additions that perhaps obscure the underlying principles.  But stacks of restricted Boltzmann machines were crucial in convincing the community that it is possible to train very deep networks.  In his Nobel Lecture Hinton referred to them as ``historical enzymes'' \cite{hinton_2024}.

Today much of the excitement surrounding artificial intelligence focuses on ``generative AI,'' such as ChatGPT and other large language models (LLMs).  A central characteristic of language is that correlations extend over very long times, so that the choice of a word is influenced by a very large context.  If we allow for arbitrary operations on this context, there is a combinatorial explosion and no reasonable amount of data will make it possible to learn what to do.  An essential ingredient in LLMs is ``attention,'' implemented in the transformer architecture, that selects which elements from the large context are relevant for particular predictions  \cite{vaswani+al_2017}.

Large language models are very far from the Hopfield model or the Boltzmann machine.  The early work of Hopfield and Hinton certainly touched off the developments that led to modern AI, in essence providing a framework within which to think very differently about brains and brain--like computation.  But there might be more.  An old problem in the Hopfield model is that it stores $K \sim \alpha N$ patterns, which can be indexed with $\sim \log N$ bits, while one might have hoped that $N$ neurons could store $\alpha N$ bits.  In 2016, Hopfield and Dmitry Krotov returned to this problem \cite{krotov+hopfield_2016}.  We can write the energy function of the original model as
\begin{equation}
E\left(  \bm{\sigma} \right)= - {J\over 2} \sum_{\rm ij} \sum_{\mu =1}^K \xi_{\rm i}^\mu \xi_{\rm j}^\mu\sigma_{\rm i}\sigma_{\rm j}  = - {J\over 2} \sum_{\mu =1}^K \left( \bm{\xi}^\mu \cdot \bm{\sigma}\right)^2 .
\end{equation}
Krotov and Hopfield explored a simple generalization,
\begin{equation}
E\left(  \bm{\sigma} \right)  = - J \sum_{\mu =1}^K F\left( \bm{\xi}^\mu \cdot \bm{\sigma}\right) .
\end{equation}
With $F(x) = x^2/2$ we have the original model, but for $F(x) = x^n$ with $n>2$ one can store $K \sim N^{n-1}$ patterns, and with $F(x) = e^x$ one can even reach  $\log K \sim \alpha N$.  These dense associative memories, or ``modern Hopfield models,'' have attracted considerable interest.  In particular, there is a connection between these models and the transformer architecture \cite{ramsauer+al_2020} and there are provocative proposals for their implementation in real brains \cite{kozachkov+al_2023}.    Krotov and Hopfield have even argued that these models with higher--order interactions can be derived from systems with pairwise interactions and hidden units, in the sprit of restricted Boltzmann machines \cite{krotov+hopfield_2021}.  Perhaps we have come full circle.

\section{Looking forward}
\label{forward}

Prediction is difficult, especially about the future,\footnote{In the twentieth century this remark was ascribed variously  to the physicist Niels Bohr (who may have adapted it from a Danish proverb), the film producer Szmuel Gelbfish (Samuel Goldwyn), the writer Samuel Clemens (Mark Twain), and the baseball player Lorenzo (Yogi) Berra.} but I have been asked to try.  I will leave aside the questions most consequential for humanity at large:\footnote{These are already the subject of a large and contentious literature.  Hinton has expressed his concerns primarily in interviews. A 2023 video provides an introduction to AI and then moves to ethical questions toward the end \cite{hinton_youtube}.} How will AI change our interactions with one another, and our sense of ourselves as human? Will we have the strength to solve the evident ethical problems that come with this new technology?  Or will we need a Hiroshima scale event to convince us that something needs to be done? Instead I will focus on questions specific to the scientific community, and especially to the biological physics community.\footnote{In this section it seems best not to give any references, lest I leave the impression that a handful of papers represent the future.}

The success of AI poses an obvious question: Why does it work?  Certainly the fact that networks with billions of parameters can learn to reach essentially optimal performance on training data means that our intuitions about optimization in high dimensional spaces were wrong.  Perhaps there is something special about neural networks that makes the parameter space navigable.  But might it then be true that the parameter space for evolutionary change also is more navigable than we thought? What about the space of parameters that describes cellular adaptation to a changing environment? It is reasonable to expect that better understanding of why artificial neural networks work will have an impact on our thinking about the physics of life.

The success of AI also answers many questions, and perhaps makes us worry that science as a professional activity may become obsolete.  In particular, the protein folding problem has long been a focus for the biological physics community, and we now have a solution:  AlphaFold, a neural network that predicts protein structures from amino acid sequences with unprecedented accuracy.  There is no question that this is a revolutionary development, not just for our field but for science quite broadly.  But there were always two parts to the protein folding problem.  The first, structure prediction, feels solved.  But the second part is why globular proteins are so different from typical heteropolymers, able to fold into well defined structures.  Given that random sequences do not fold, can we define an order parameter that measures the ``protein--ness'' of an arbitrary sequence?  More generally, we can hope that the success of AI in answering practical questions will focus our attention on deeper conceptual issues.

The success of AI enables us to do new science, much as with other tools.  To give just one example, there has been a quiet revolution in the ``physics of behavior.''  Large scale, high resolution data on animal movements in increasingly natural contexts has made it possible to discover underlying regularities, combining the precision of psychophysics with the complexity of ethology.  The first steps in these analyses involve reducing raw video to a set of variables whose trajectories capture the behavior.  Just a decade ago this required hand tuned methods, specific to each organism or context.  Today we can use neural networks as a shortcut, learning from a limited training set how to extract meaningful trajectories from video.  Such data reduction is not a theory of the underlying behavior, but doing the reduction more efficiently allows for much wider exploration.  Again, solving practical problems should focus our attention on deeper conceptual issues.

An essential feature of artificial neural networks is that it is nearly impossible to point to a particular node in the network and ``say what it does.''  Perhaps this is an inevitable consequence of computational functions being emergent.  If the functional behaviors of living systems are emergent in this sense, then there are limits on our ability to point to component parts and describe their contributions to the performance of the system.  There is new physics to be uncovered in the relation between microscopic mechanisms and macroscopic functions.

The current excitement around AI can be consuming.  I have tried to emphasize that at least part of how we got here is through explorations in the physics of life.  We might have had ChatGPT without Boltzmann machines or the Hopfield model, but the actual path did go through these ideas.  And these ideas emerged after Hopfield had been thinking for a decade about the physics problems behind many of life's phenomena---allostery, the transfer of electrons between proteins, the reliability of information transmission by copying of molecules.  The theme that runs through this work and continues in the development of neural network models is the belief that there {\em are} general physical principles to be found.  To get to these principles we might need to start by ignoring details, but success is when the resulting theory serves to organize these details, as with the seemingly wasteful side branches in otherwise linear pathways for DNA replication, transcription, and protein synthesis.   

When Hopfield began to explore living systems, fifty years ago, there was a widespread prejudice among physicists that these systems were too complicated and too messy to yield to the physicist's style of inquiry.  John made wise choices, looking at problems where experimentalists had tamed the complexity to the point where they could do experiments with the precision and reproducibility that we expect in physics.  At that time, this meant focusing largely on the behavior of isolated molecules.  

In the five decades since Hopfield’s original work on hemoglobin, a handful of protein structures became the protein data bank and this tremendous body of data made it possible for AlphaFold to learn the rules of protein structure prediction.  Hopfield’s focus on collective coordinates in proteins continues to resonate, most recently in unifying the functional dynamics of synapses across many decades in time. Beyond monitoring the accuracy with which molecules are copied or synthesized, it now is possible to monitor the precision with which cells control and sense concentrations of crucial signaling molecules, testing candidate physical principles that may govern information flow through the underlying networks.  The ability to record the electrical activity of hundreds or thousands of neurons simultaneously creates opportunities to bring statistical physics ideas---Hopfield models, Boltzmann machines, and more---into direct contact with data on real brains, revealing new emergent phenomena. As noted above, the classical problems of animal behavior have been reinvigorated by AI--enhanced tools to make quantitative measurements under increasingly natural conditions, while explorations of flocks and swarms have uncovered new universality classes.  

The level of experimental control needed to ``do physics'' in the complex context of living systems now reaches broadly across scales, from molecules to cells all the way to ecology and collective behaviors in animal groups.   Today's theorists thus have a much larger playground to explore.  With luck, we will see more grand successes, in the spirit of Hopfield's remarkable work.

\section{But is it physics?}
\label{isitphy}

At a press conference on the afternoon of the Nobel Prize announcement, a reporter from the {\em Washington Post} submitted a question for John: ``Some people are wondering how the work awarded today, which has revolutionized computer science, fits within the field of physics---what would you say to them?'' She might have asked more simply ``but is it physics?''

Many of us in the biological physics community have heard this question, sometimes in harsh terms.  It seems worth noting that Nobel committee addressed this question directly, stating explicitly that ``with artificial neural networks the boundaries of physics are extended to host phenomena of life as well as computation'' \cite{nobel_website}.

Although best left unreferenced, even some professional physicists took to social media to denounce the prize as not being for real physics.  The recognition of AlphaFold   \cite{jumper+al_2021} by the Nobel Prize in Chemistry the next day led to claims that our colleagues in Stockholm had succumbed to AI hype.  It was even suggested that the next logical step would be to award the Nobel Prize in Literature to ChatGPT. 

These arguments about whether progress belongs to one discipline or another implicitly assume that the boundaries of disciplines are fixed.  In the 1960s there was skepticism about whether looking at DNA was ``real biology.''   General relativity is a crowning achievement of theoretical physics, but decades with no connections to experiment meant it was pursued more in mathematics departments than in physics departments.   Cosmology went from a subject of religious speculation to a core topic in physics over the course of a lifetime, and much of what we now describe as quantum information might once have been categorized as the philosophical foundations of quantum mechanics.  It took a long time to convince the whole physics community that there was anything fundamental to be discovered by studying solids; the field was derided as ``squalid state physics'' and applied physics departments were formed to be sure that it did not overwhelm ``real physics.''   

These examples, and many others, emphasize that the boundaries of disciplines are not static.  When exciting things happen at the borders between disciplines, that excitement need not remain confined at the interface.  New fields can break off, and the definitions of established fields can change.  For physics especially, there is a strong argument that what defines the discipline  is not the objects that we study but rather the style of inquiry, the kinds of questions that we ask and the kinds of answers that we seek \cite{decadal_survey,bialek_2012}. It is not that everything is physics, but rather that physicists ask new questions in contexts that have been explored by practitioners of other disciplines, and that these problems become part of what we recognize as Physics (with a capital P).  It is  best to give the last words on these issues to Hopfield himself \cite{hopfield_2014}:

``What is physics? To me---growing up with a father and mother who were both physicists---physics was not subject matter. The atom, the troposphere, the nucleus, a piece of glass, the washing machine, my bicycle, the phonograph, a magnet---these were all incidentally the subject matter. The central idea was that the world is understandable, that you should be able to take anything apart, understand the relationships between its constituents, do experiments, and on that basis be able to develop a quantitative understanding of its behavior. Physics was a point of view that the world around us is, with effort, ingenuity, and adequate resources, understandable in a predictive and reasonably quantitative fashion. Being a physicist is a dedication to the quest for this kind of understanding.''

\begin{acknowledgements}
It is a pleasure to thank John Hopfield for teaching me so much about so many things.  I feel very fortunate that the somewhat terrifying figure whom I met when I was a young student became a mentor, colleague, and friend.  My thanks also to colleagues who gave advice on the manuscript.  Errors that remain are mine alone.
\end{acknowledgements}

\end{document}